\documentclass[aps,prb,twocolumn, superscriptaddress]{revtex4-1}
\usepackage{amsmath}
\usepackage{amssymb}
\usepackage{graphicx}

\begin{document}

\title{Chemical mechanical polishing of thin film diamond}

\author{Evan L. H. Thomas}
\email[Corresponding Author:]{ThomasEL10@Cardiff.ac.uk}
\affiliation{School of Physics and Astronomy, Cardiff University, Cardiff, UK}
\author{Geoffrey W. Nelson}
\affiliation{Department of Materials, Imperial College, London, UK}
\author{Soumen Mandal}
\affiliation{School of Physics and Astronomy, Cardiff University, Cardiff, UK}
\author{John S. Foord}
\affiliation{Department of Chemistry, University of Oxford, Oxford, UK}
\author{Oliver A. Williams}
\email[Corresponding Author:]{WilliamsO@Cardiff.ac.uk}
\affiliation{School of Physics and Astronomy, Cardiff University, Cardiff, UK}

\date{\today}

\begin{abstract}
The demonstration that Nanocrystalline Diamond (NCD) can retain the superior Young's modulus (1,100 GPa) of single crystal diamond twinned with its ability to be grown at low temperatures ($<$450 $^o$C) has driven a revival into the growth and applications of NCD thin films.  However, owing to the competitive growth of crystals the resulting film has a roughness that evolves with film thickness, preventing NCD films from reaching their full potential in devices where a smooth film is required.  To reduce this roughness, films have been polished using Chemical Mechanical Polishing.  A Logitech Tribo CMP tool equipped with a polyurethane/polyester polishing cloth and an alkaline colloidal silica polishing fluid has been used to polish NCD films.  The resulting films have been characterised with Atomic Force Microscopy, Scanning Electron Microscopy and X-ray Photoelectron Spectroscopy.  Root mean square roughness values have been reduced from 18.3 nm to 1.7 nm over 25 $\mu$m$^2$, with roughness values as low as 0.42 nm over $\sim$0.25 $\mu$m$^2$.  A polishing mechanism of wet oxidation of the surface, attachment of silica particles and subsequent shearing away of carbon has also been proposed.\end{abstract}

\maketitle

\section{Introduction}
The demonstration that Nanocrystalline Diamond (NCD) retains many of the superlative properties of single crystal diamond in a low cost, large area wafer scale package, as well as the possibility of CMOS integration due to the possibility of growth at low temperatures ($<$450 $^o$C) has driven a resurgence in research into the use of thin diamond films \cite{WilliamsND, CVDbutler}.  With a high Young's modulus of 1,100 GPa, the highest phase velocity of all materials of 12,000 m/s and thermal conductivity up to 2,000 W/mK \cite{philipjap}, applications include Micro-Electro-Mechanical Systems (MEMS), Surface Acoustic Wave (SAW) devices, thermal management and tribological coatings.  

However, diamond does not grow epitaxially on silicon, requiring wafers to be seeded with nanodiamond particles prior to growth \cite{williamscpl}. The subsequent competitive growth of these nanocrystals into coalesced NCD films results in a surface roughness that evolves with film thickness and exceeds that of cleaved single crystal diamond.  The increased roughness of NCD films can be detrimental for many of its key applications, such as the integration of AlN as a piezoelectric in MEMS and SAW applications, decreased Q - spoiling of MEMS devices and enhanced friction in tribological coatings.

To work around this roughness, previous reports have used the nucleation side of freestanding NCD films, either locally removing the silicon substrate or bonding / glueing the wafer to another support and completely removing the silicon \cite{rodriguezsaa}. However, this process is complicated, time consuming and incompatible with some applications of NCD such as tribology and most MEMS structures. Another approach is to interrupt the crystallite expansion by a re-nucleation process that results in smaller crystallite sizes and reduced surface roughness \cite{jiaojap, haubnerijrm}. However, this process has been shown to result in reduced Young’s modulus and optical transparency due to the greater contents of sp$^2$ carbon in the resulting films \cite{haubnerijrm, williamscpl2010}. Therefore, there is a real need for a polishing step to produce low roughness NCD films.

\begin{figure}
\includegraphics[width=7cm]{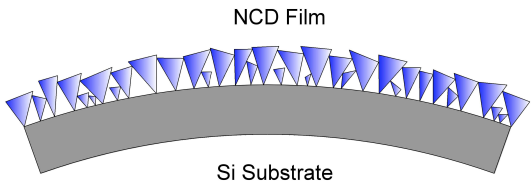}%
\caption{Schematic of exaggerated wafer bow seen with NCD films. Due to the differing coefficients of thermal expansion of diamond and silicon, upon cooling from growth temperatures significant bowing will occur.\label{fig1}}
\end{figure}

Traditional mechanical polishing of diamond involves pressing a sample against a fast rotating iron scaife,  $>$ 2,500 rpm, in the presence of a diamond grit and binder.  With forces greater than 10 N micro-cracking of the diamond occurs, with roughness values dependent on the grade of grit used \cite{malshedrm, schuelkedrm}.  However with this technique removal rates are generally low, $\sim$10 nm/hr, and the high forces on the sample can cause deep fissures and create surface pits \cite{schuelkedrm}.  To enhance the polishing rate and reduce surface damage the hybrid technique of Chemically Assisted Mechanical Polishing and Planarisation (CAMPP) was developed.  With this technique an oxidiser, typically potassium nitrate or potassium hydroxide, heated to around 360 $^o$C is added to the mechanical process \cite{malshedrm, ollisondrm}.  After cracking by the scaife, the molten oxidisers enter and convert diamond to carbon dioxide and carbon monoxide weakening the surface and allowing further micro-cracking to occur.  Through this technique faster removal rates and arithmetic roughness (R$_a$) values of 2.8 nm are achievable when used in conjunction with an initial mechanical polish \cite{ollisondrm}. However, while this technique makes it possible to polish films of several tens of microns thickness, for films with thickness in the hundreds of nanometres the wafer bow can be significantly greater than the thickness of the film (typically $>$10$\mu$m over a 2'' Si wafer), as shown schematically in figure \ref{fig1}. This will prevent uniform polishing across the entire film and possibly cause shattering due to the rigidity of the scaife.  Therefore a more flexible polishing pad is required in order to conform to the bowed sample.  One such method that is commonly used in the IC fabrication industry for the polishing of dielectric and metal interconnects is Chemical Mechanical Polishing (CMP).  With this technique a softer polyester based polishing pad is used rather than a hard metal scaife with the aid of a colloidal silica (50-200 nm) \cite{zantyemse} based polishing slurry at room temperatures.  In conventional dielectric polishing Silicon Dioxide (SiO$_2$) is converted to bound silanol groups (Si(OH)$_4$) by the liquid polishing fluid, silica particles in the slurry then bond to the surface of passivation layer \cite{krishnancr}.  The moving polishing pad, if sufficiently rough, will then create a force on the silica particle.  As long as the shear force applied is larger than the binding energy, the polishing pad then removes the particle and attached silanol molecule from the surface \cite{hochengjtes}.

In this paper, CMP of NCD films is reported with the use of silica based polishing fluid and a polyester/polyurethane polishing pad at room temperature.  It is important to note that no diamond-based products were used in either the pad or slurry, unlike previous studies \cite{hochengbook}.  Films have been studied with Scanning Electron Microscopy (SEM) to deduce morphology, Atomic Force Microscopy (AFM) to deduce roughness, and X-ray Photoelectron Spectroscopy (XPS) in an effort to explain the polishing mechanism.

\section{Experimental Procedure}
Silicon (100) p-type 2-inch wafers of 500 $\mu$m thickness were used as substrates throughout. Before deposition all wafers were cleaned using the standard SC-1 process of 30\% H$_2$O$_2$ : NH$_4$OH : DI H$_2$O (1:1:5) at 75 $^o$C for 10 minutes. The substrates were then rinsed in DI H$_2$O in an ultrasonic bath for 10 minutes and spun dry. To seed, the wafers were placed in a mono-dispersed nanodiamond/H$_2$O colloid and agitated in an ultrasonic bath for 10 minutes. This process is known to produce nucleation densities exceeding 10$^{11}$ cm$^{-2}$. \cite{williamscpl} Once seeded the wafers were rinsed, spun dry at 3,000 rpm, and then immediately placed inside the CVD chamber.

Chemical Vapour Deposition (CVD) of Nanocrystalline Diamond was carried out in a Seki 6500 series Microwave Plasma Reactor under 3\% CH$_4$/H$_2$ conditions at 40 Torr and 3.5 kW microwave power. Upon termination of growth all films were cooled down in hydrogen plasma to ensure hydrogen termination and prevent deposition of non-sp$^3$ material. Substrate temperatures were approximately 840 $^o$C as determined by dual wavelength pyrometry, with heating solely from the microwave induced plasma. Films were grown to 360 nm determined in-situ through the use of pyrometric interferometry, and ex-situ with a Filmetrics F-20 Spectral Reflectance system.

\begin{figure}
\includegraphics[width=7cm]{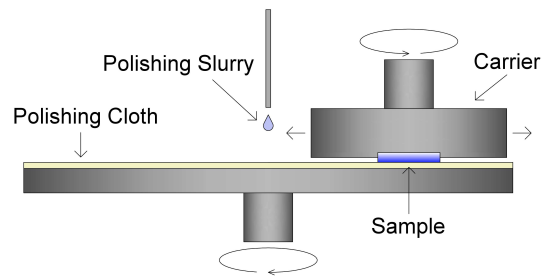}%
\caption{Schematic diagram of CMP tool.  Sample held in rotating carrier and swept across counter rotating polishing cloth.  Slurry distributed continually onto top of polishing cloth.\label{fig2}}
\end{figure}

Chemical Mechanical Polishing was performed with a Logitech Tribo polishing system in conjunction with a SUBA-X polishing pad and a Logitech supplied alkaline colloidal silica polishing slurry (Syton SF-1). Before use, the pad was conditioned for 30 minutes to ensure a high surface roughness to maximise polishing action and slurry distribution \cite{mcgrathjmpt}. During polishing both pad and carrier were kept at 60 rpm rotating in opposite directions, while the carrier swept across the pad as shown schematically in figure \ref{fig2}. Down pressure was kept at 4 psi, while a backing pressure of 20 psi was used in an attempt to present a flat NCD film surface to the polishing pad. After initial wetting of the plate, the feed slurry rate was kept at 40 ml/min. three films were polished for durations of 1, 2, and 4 hours. After polishing the films were cleaned in an attempt to remove any remaining polishing slurry with a standard SC-1 clean as detailed previously.

SEM images were taken with the SE2 detector of a Raith e-line SEM, operated at 10 kV, 10 mm working distance and 20 k magnification. AFM was performed with a Veeco Dimension 3100 AFM operated in tapping mode and equipped with a TESPA tip of 320 kHz resonant frequency, 8 nm radius, and 42 N/m spring constant. 5 areas of 25 $\mu$m$^2$ were taken around the centre of each sample, with post AFM analysis being carried out with Gwyddion SPM analysis software.  Removal rates were calculated by comparing the average thickness of 13 points on each film before and after polishing with the Filmetrics F-20 system. 

XPS experiments were conducted using a VG ESCA Lab XPS spectrometer at 1 $\times$ 10$^{-9}$ Torr,  using an Al K$\alpha$ radiation source (1486.3 eV) at 10 kV anode with 10 mA emission current.  The Fixed Analyser Transmission (FAT) mode was used to obtain spectra, using a pass energy of 50 eV or 25 eV for survey and ‘narrow’ XPS scans, respectively.  All peak fitting was done using XPS Peak Fit (v. 4.1) software.  The reported binding energies have an error of $\pm$ 0.25 eV, based on the calibration to the C1s peak.  Peak areas were normalized to the XPS cross-section of the F1s photoelectron signal by use of the atomic sensitivity factors \cite{wagnersia}.  Elemental ratios were calculated from the normalized peak areas and have an error of about 15-20\% \cite{saa}.

\section{Results and Discussion}
\subsection{Morphology}
\begin{figure}
\includegraphics[width=8cm]{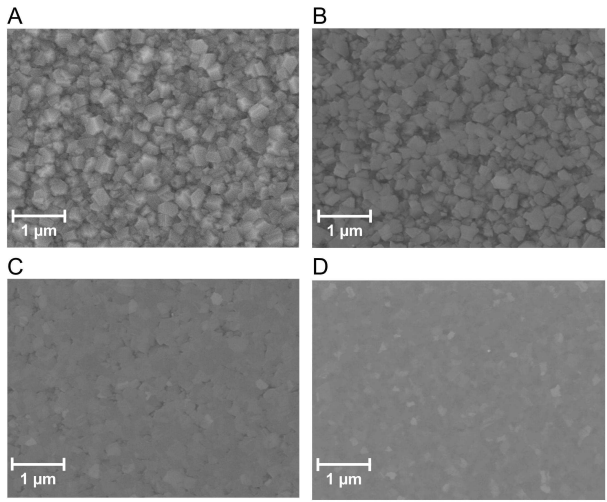}%
\caption{SEM micrographs of as grown and polished films. CMP was used on three different films for the indicated amount of time under identical conditions. A) As grown, B) 1 hour CMP film, C) 2 hours, and D) 4 hours. \label{fig3}}
\end{figure}
SEM images of the as grown and 1-4 hour polished films are shown in figure \ref{fig3}.  The as grown film of figure \ref{fig3}A show clear faceting with crystal sizes of approximately 100 - 250 nm, as is typical for films grown under 3\% methane admixture conditions \cite{williamscpl2010}.  When comparing the as grown film to the 1 hour CMP film of figure \ref{fig3}B a clear polishing action can be seen.  Peaks of the crystals that come into contact with the polishing pad are removed first, followed by a progression down to the point at which a neighboring crystal is met for the 2 hour and 4 hour films of figures \ref{fig3}C and D.  The resulting crystal plateaus appear very smooth with little evidence of cracking, suggesting a significant chemical action to the polishing.  After 4 hours of polishing it can be seen that the film appears close to optimum.  With reference to figure \ref{fig2} it can also be seen that a point has been reached at which the majority of crystal peaks are removed, while there is also a lack of voids opening up to the substrate.  Very little contamination from silica can also be seen on the SEM images, initially suggesting an SC-1 clean is enough to clear the surface of any loose polishing introduced contamination.

\begin{figure}
\includegraphics[width=8cm]{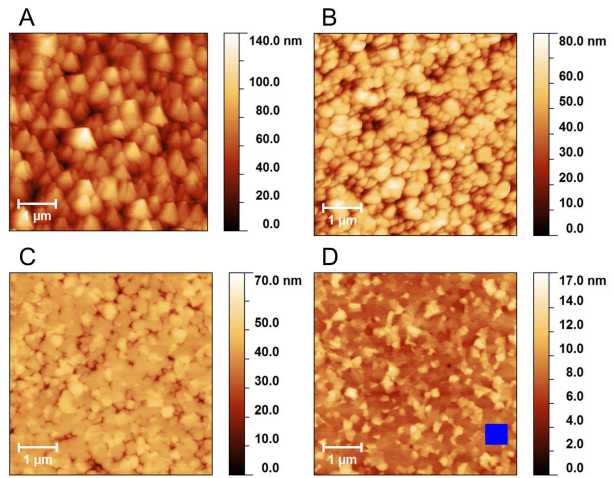}%
\caption{Corresponding AFM micrographs for as grown and polished films shown in figure \ref{fig3}.  A) As grown, B) 1 hour CMP film, C) 2 hours, and D) 4 hours. \label{fig4}}
\end{figure}
Figure \ref{fig4} shows the AFM images of the as grown and 1-4 hour CMP films, while Table \ref{tab1} shows the average roughness over the 5 scans of 25 $\mu$m$^2$ for each film.  As can be seen the micrographs reiterate this steady polishing showing a decrease in roughness from the as grown 18.3 nm rms to 1.7 nm rms over the 25 $\mu$m$^2$ scans.  Also shown in blue on figure \ref{fig4}D is a smaller area of 0.25 $\mu$m$^2$ showing that a local roughness of 0.42 nm rms is achievable with CMP and the parameters used.  The removal rate is approximately 16 nm/hr for the three polished films. 

\begin{table}
\caption{Roughness values over 25 $\mu$m$^2$ for as grown and polished films. \label{tab1}}
\begin{tabular}{c|c}
\hline 
\textbf{Polishing Duration (hrs.)} & \textbf{Roughness (nm rms)} \\ \hline 
0 & 18.3 \\
1 & 11 \\
2 & 4.5 \\
4 & 1.7 \\ \hline
\end{tabular}
\end{table}

As attempts to use Raman and the surface enhanced technique of Shell Isolated Nanoparticle Enhanced Raman Spectroscopy (SHINERS) \cite{linature} were deemed inconclusive due to the swamping of the surface signal by the signal from the bulk, XPS has been used to deduce polishing mechanism.

\subsection{X-ray Photoelectron Spectroscopy}
\begin{figure*}
\includegraphics[width=17cm]{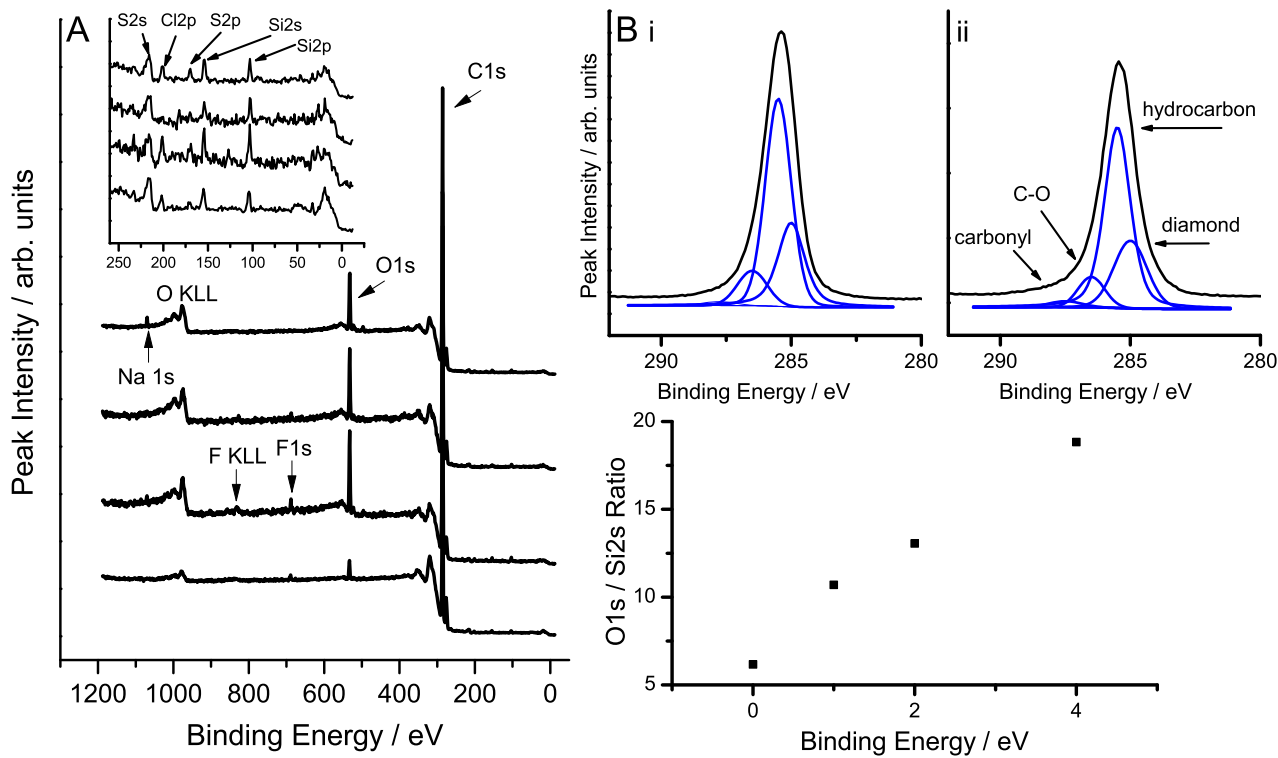}%
\caption{A) Survey XPS spectra of diamond substrates with main photoelectron and Auger signals
indicated. From top to bottom, the spectra represent as grown, 1 hour CMP, 2, and 4 hours
respectively. The insert contains data from the main body of the figure, showing XPS features at low
binding energy. B) Representative C1s spectra before i) and after 4 hours ii) of polishing. The experimental
data (black) is shown above the deconvoluted signal (blue). C) O1s/Si2s ratio plotted as a function of the
polishing duration.\label{fig5}}
\end{figure*}
XPS has been widely used to study CVD diamond films and previous studies are the basis for the present analysis \cite{ferrocarbon, wilsonjesr}.  Survey XPS spectra are shown in figure \ref{fig5}A, with major photoelectron and Auger peaks labelled.  Unpolished and polished diamond films have significant C1s ($\approx$285.0 eV) and O1s ($\approx$531.0 eV) character.  Photoelectron signal originating from F, S, Cl and Si core levels are seen, particularly on polished diamond films.  Clearly the chemical polishing process is introducing non-diamond contamination to the surface.  However, this level of contamination is not sufficient enough to be detrimental to the use of NCD in MEMS devices and the applications mentioned earlier. It is also highly probable that similar post CMP cleaning processes to the CMOS industry can be developed to remove this surface contamination such as hydrogen or oxygen plasma exposure.   

The C1s region is typically used to characterise changes to the surface chemistry of diamond thin films.  Representative C1s spectra are shown in Fig. 5B.  The C1s spectra were deconvoluted into four chemical environments, as done for previous studies on CVD diamond thin films \cite{ferrocarbon, wilsonjesr}: diamond  (C-C, 285.0 eV), hydrocarbon (C-H, 285.5 eV), ether (C-O, 286.5 eV), and carbonyl (C=O, 287.5 eV).  No evidence for the presence of a significant amount of a fifth form, carboxyl (C(=O)OH, 288.5 eV) was found.  By comparing figure \ref{fig5}Bi and Bii, it can be seen that CMP does not significantly change the chemical termination of the CVD diamond surface, although  it does lead to subtle changes in the concentrations of the differing carbon species.  However the most important conclusion from the chemical analysis is that significant amounts of graphite or graphite related defects do not develop on the surface:  it is well-known that treatments such as Ar ion bombardment \cite{laujap} and electrochemical anodisation \cite{foordpssa}  produce an sp$^2$ type defect structure on the diamond interface which is visible in XPS as a peak shifted by about 1 eV to lower binding energy of the main diamond peak.  It can be seen from figure \ref{fig5}Bii that such defects are not produced by the polishing procedure, again emphasising the gentle nature of CMP.

\begin{figure}
\includegraphics[width=8cm]{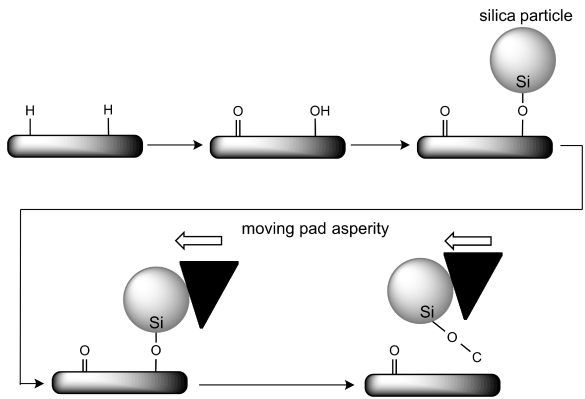}%
\caption{Proposed polishing mechanism.  Wet oxidation of hydrogen terminated diamond by polishing fluid increases the carboxyl (COOH), carbonyl (C=O), and hydroxide (OH) content on the surface.  As with the CMP of SiO$_2$, hydroxide ions facilitate attachment of silica particles to surface.  Shearing forces generated on silica particle by asperities of the rough polishing pad then removes carbon atom from surface, providing polishing. \label{fig6}}
\end{figure}

\begin{table}
\caption{Elemental ratios of O1s, F1s, and Si2s with respect to C1s as a function of polishing duration. \label{tab2}}
\begin{tabular}{c|c|c|c}
\hline 
\textbf{Polishing Duration} & \textbf{O1s/C1s} & \textbf{F1s/C1s} & \textbf{Si2s/C1s} \\ 
\textbf{(hrs.)} &&& \\ \hline 
0 & 0.022 & 0.017 & 0.004 \\
1 & 0.142 & 0.076 & 0.013 \\
2 & 0.120 & 0.032 & 0.009 \\
4 & 0.100 & 0.010 & 0.005 \\ \hline
\end{tabular}
\end{table}

The polishing process permanently increases the oxygen character of the diamond surface.  This is clearly demonstrated by the dramatic increase in the O1s/C1s ratio after 1 hour of polishing (see Table \ref{tab2}).   Given that the XPS sampling depth in diamond is probably up to 10 C layers, based on typical electron elastic mean free paths, O:C ratios of 0.1 signify at least monolayer coverage.  Curiously however the level of C oxidation does not seem to be changed by this increase in oxygen. However noting the presence of the addtional elements F, Cl, Si \& S,  it is clear the associated molecular species  could contain O, and of course adsorbed water could also be present. The  trend in the O1s/Si2s ratio in figure \ref{fig5}C shows this increases with polishing time.  This emphasises the point that the source of O in the XPS spectra is not limited to silica, but has a main component from other species. Overall the conclusion is that a range of molecular species which are fairly strongly bound (and therefore surviving the cleaning of the samples  before XPS analysis) do form on the surface being polished.

The F1s signal after 1 hour of polishing was unexpected.   After 4 hours, this signal is reduced to near-negligible levels (see Table \ref{tab2}).   The origin of this signal is likely the polymer-based pads used to polish the diamond substrates, surfactants in the chemical solution, or solvent residue.  All of these sources could be additional sources of Cl, S, \& O photoelectron signal. 

\section{Discussion}
The SEM and AFM images of figure \ref{fig4} and \ref{fig5} show a steady polishing action with time.  The polishing begins with the removal of peaks due to the contact with polishing pad, followed by a progression down to the intersection with neighbouring crystals.  This initial polishing of high points with the smoothness and apparent crack free nature of crystal tops suggests a true chemical and mechanical synergy to the polishing.  This steady polishing is reiterated by the AFM with an as grown rms roughness of 18.3 nm being reduced to 1.7 nm after 4 hours.  It can also be seen that for a smaller area, closer to the size of an individual crystal, the roughness can be as low as 0.42 nm.  The removal rate seen of approximately 16 nm/hr exceeds that typically possible with traditional mechanical polishing \cite{schuelkedrm}, however this is heavily dependant on the age and condition of the pad.  While this is less than the $\mu$m/hr polishing rates of CAMPP, films in the hundreds of nanometres with lower initial roughness can be polished without possible cracking.

With regards to mechanism, contact polishing can be broadly divided into three mechanisms: micro-chipping, conversion to graphite, and chemical reaction\cite{malshedrm}.  Due to the comparatively low hardness and flexibility of the polyester/polyurethane polishing cloth as well as the lack of any diamond based products in both the slurry and cloth, it is unlikely that micro-chipping is the cause of polishing.  Coupled with the low temperatures used and the lack of strong oxidizing agent it is also unlikely that conversion to CO and CO$_2$ can be responsible with the polishing rates seen.

The lack of significant change in the graphitic content of the films as indicated by XPS indicates that conversion to graphite is not responsible for polishing.  Typical techniques that rely on conversion to graphite utilise catalytic materials such as iron cobalt or nickel to lower the activation energy and operate at temperatures of approximately 750 $^o$C\cite{malshedrm}, significantly higher then the 30-50 $^o$C temperature of the waste slurry.  

Therefore it is proposed that the polishing mechanism follows that of the CMP of Silicon Dioxide.  In traditional CMP, hydroxide ions within the polishing fluid react with the surface siloxane (Si-O-Si) bonds, creating a silanol based passivation layer (Si(OH)$_4$)\cite{krishnancr, hochengjtes}.  Silica particles within the polishing fluid will then attach themselves to the hydrated groups of the passivation layer.  Should the polishing pad then be sufficiently rough, a shearing force will be created on the silica particles.  If this force is larger than the binding energy, the molecule will be removed resulting in polishing of the surface.

With diamond XPS has shown that CMP leads to general oxidation of the interfacial region; increasing the carbonyl and hydroxyl content of the surface.  Drawing parallels to the hydroxyl bonding seen in the polishing of SiO$_2$, we believe the OH termination facilitates the bonding of silica particles to the surface, as shown schematically in figure \ref{fig6}.  As with SiO$_2$ CMP the rough pad surface will then create a shear force on the silica particle.  Due to the bond strengths of Si-O, O-C and C-C being 800 kJ/mol, 1077 kJ/mol and 610 kJ/mol respectively \cite{crc}, it is believed that when this force is applied the C-C bond will break, polishing the film surface.  Alternatively an oxidised silica particle can directly attach itself without the need for intermediate wet chemical oxidation.  As this is only a proposed model based on the mechanism on SiO$_2$, further work is needed for validation and optimisation of the CMP of diamond films.

Through the use of CMP it has been shown that bowed thin film diamond can be polished without fear of cracking of films.  The technique removes the need for the use of expensive diamond grit, or cast iron scaifes and instead uses polyester/polyurethane polishing pads commonly found in the IC fabrication industry.  As shown, considerable action can be seen without the need for raised temperatures or high pressures, simplifying equipment required.  Therefore CMP is a promising method of achieving low roughness diamond surfaces at low cost.

\section{Conclusion}
NCD films have been polished by CMP with the use of a polyurethane/polyester felt and an alkaline colloidal silica polishing fluid (Syton SF-1).  No diamond based products were used in either the slurry or polishing cloth. Final roughness values of 1.7 nm rms were achieved over 25 $\mu$m$^2$, with values as low as 0.42 nm rms over $\sim$0.25 $\mu$m$^2$.  The polishing mechanism proposed consists of the wet oxidation of the surfaces with the polishing fluid facilitating the attachment of silica particles to the diamond film, followed by the shearing away of the particle due to forces from the polishing pad.  Thus with its low temperature, simple operation, ability to polish wafers with significant bow and already common CMOS industry supplies, CMP is an attractive method for the polishing of thin film diamond.

\begin{acknowledgments}
 The authors acknowledge the financial support of the EPSRC under the grant ‘Nanocrystalline diamond for
Micro-Electro-Mechanical-Systems’ reference number EP/J009814/1. They also wish to thank Sam Ladak
and Dan Read at the Cardiff School of Physics and Astronomy for assistance with AFM.
\end{acknowledgments}

\bibliography{ref}

\end{document}